\documentclass[preprint,authoryear,12pt]{elsarticle}

\usepackage{amssymb}
\usepackage{pdflscape}

\makeatletter
\def\ps@pprintTitle{%
  \let\@oddhead\@empty
  \let\@evenhead\@empty
  \let\@oddfoot\@empty
  \let\@evenfoot\@oddfoot
}
\makeatother
\begin{document}
\begin{frontmatter}



\title{Is there a housing bubble in China?}


\author[1]{Tianhao Zhi}
\author[1]{Zhongfei Li}
\author[2]{Zhiqiang Jiang}
\author[1]{Lijian Wei \corref{cor1}}
\author[3]{Didier Sornette}
\address[1]{Business School, Sun Yat-sen University, Guangzhou 510275, China}
\address[2]{Department of Finance, East China University of Science and Technology, Shanghai 200237,  China}
\address[3]{Chair of Entrepreneurial Risks and The Financial Crisis Observatory, Department of Management, Technology and Economics, Scheuchzerstrasse 7, CH-8092 Zurich, ETH Zurich, Switzerland}
\cortext[cor1]{Corresponding author: Lijian Wei. Email: weilj5@mail.sysu.edu.cn}

\begin{abstract}
There is a growing concern in recent years over the potential formation of bubbles in the Chinese real estate market. This paper aims to conduct a series of bubble diagnostic analysis over nine representative Chinese cities from two aspects. First, we investigate whether the prices had been significantly deviating from economic fundamentals by applying a standard Engle-Granger cointegration test. Second, we apply the Log-Periodic-Power-Law-Singularity (LPPLS) model to detect whether there is any evidence of unsustainable, self-reinforcing speculative behaviours amongst the price series. We propose that, given the heterogeneity that exists amongst cities with different types of bubble signatures, it is vital to conduct bubble diagnostic tests and implement relevant policies toward specific bubble characteristics, rather than enforcing one-that-fits-for-all type policy that does not take into account such heterogeneity.
\end{abstract}

\begin{keyword}
housing bubble, Log-Period-Power-Law-Singularity, bubble prediction, the Chinese real estate market.
\\
\emph{JEL classification}:  F37, G01, G17
\end{keyword}

\end{frontmatter}


\newpage
\section{Introduction}
\label{intro}
\begin{figure}[h!]
  \centering
  \includegraphics[scale=0.9]{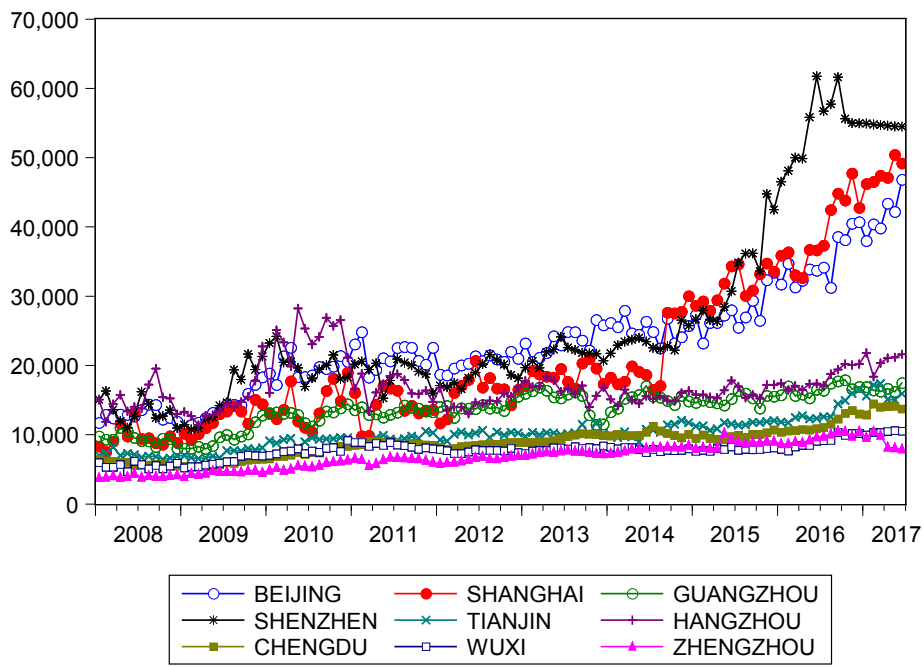}\\
  \caption{Residential real estate prices ($Yuan/m^2$) in China from Jan/2008 to Jun/2017. There is a significant rise in most Tier-1 cities, while prices of Tier-2 and Tier-3 cities had grown more moderately. \emph{Source: }WIND Database}
  \label{fig:price}
\end{figure}

The Chinese housing market had experienced an unprecedented boom since 2008. According to Fig. \ref{fig:price}, the price in Shenzhen had nearly quadrupled from Jan/2008 to Jun/2017. Beijing and Shanghai had grown by nearly three times over the recent decade. Some cities outside the Tier-1 category also had a remarkable growth. While some analysts were concerned, posing dire warnings of the imminent collapse of the Chinese housing market, others reject the bubble hypothesis, finding justifications of the booming real estate market in terms of the rapid pace of urbanization and other social factors such as the cultural urge to hold real estates \citep{ACHPZ2010}.

\begin{figure}[h!]
  \centering
  \caption{Top panel: Growth rate of GDP as a function time from 2005 to 2017; bottom panel: Li Keqiang index as defined in footnote 2. The GDP growth had steadily fallen over the decade, accompanied by similar downward trends in most sub-sectors. The finance sector had boomed over a brief period since 2014, followed by a sharp decrease in mid-2015 due to the stock market crash over that period. See \citet{JZSWBC2010} for detailed analysis over the stock market crash. \emph{Source:} WIND Database}
  \begin{tabular}{cc}
    \includegraphics[scale=0.65]{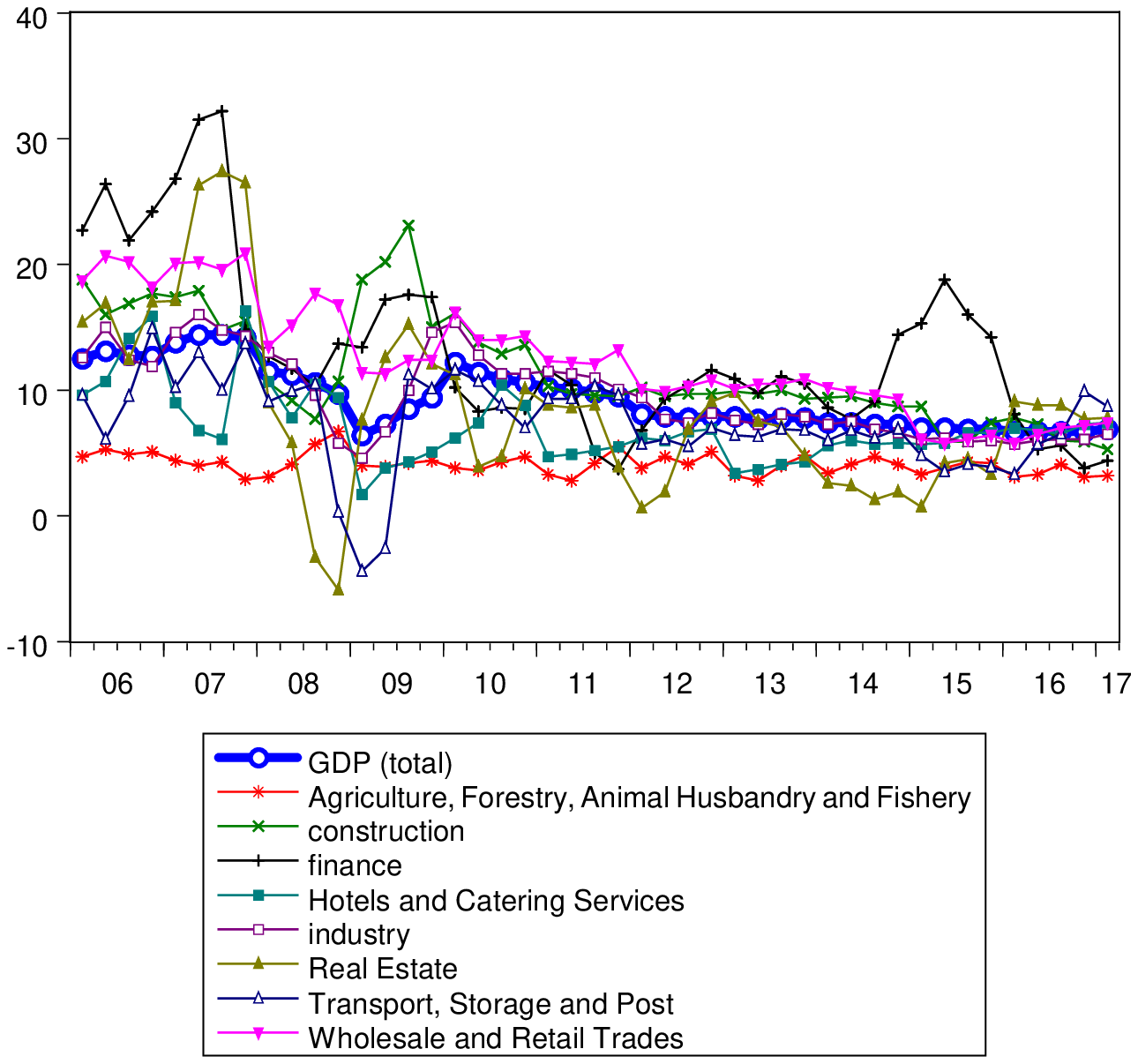}\\
    \includegraphics[scale=0.65]{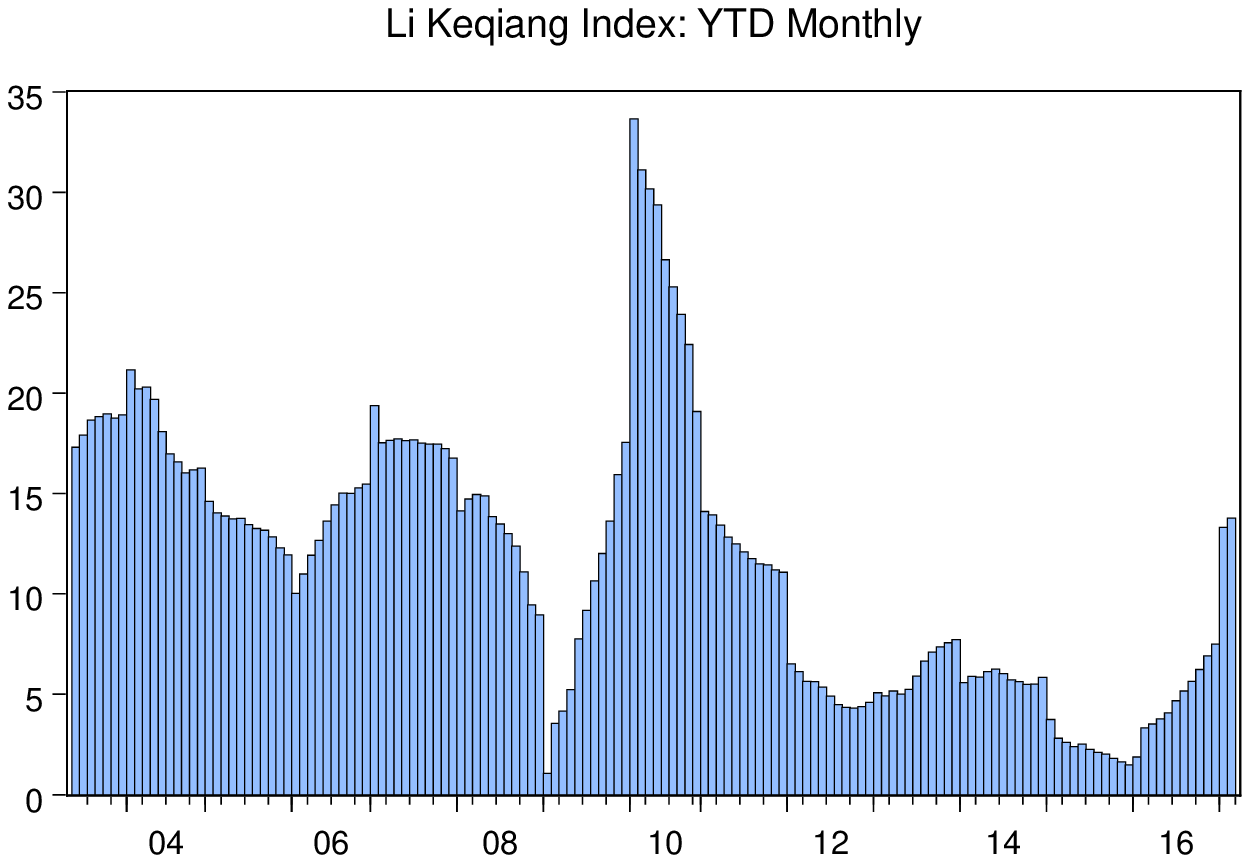}
  \end{tabular}
    \label{fig:GDP}
\end{figure}

\begin{figure}[h!]
  \centering
  \caption{Top-left panel: Provident Fund Interest Rate as a function of time. Top-right panel: Stock of New Medium \& Long-Term RMB loans to Households (\emph{CNY 100mn}) Bottom-left panel: Elasticity of Personal Housing Loan (PHL), Real Estate Development Loan (RED), and Commercial Real-estate Loan (CRE) relative to GDP. Bottom-right panel: Proportion of Personal Housing Loan (PHL) over total assets in Big-4 Banks}
  \begin{tabular}{cc}
    \includegraphics[scale=0.45]{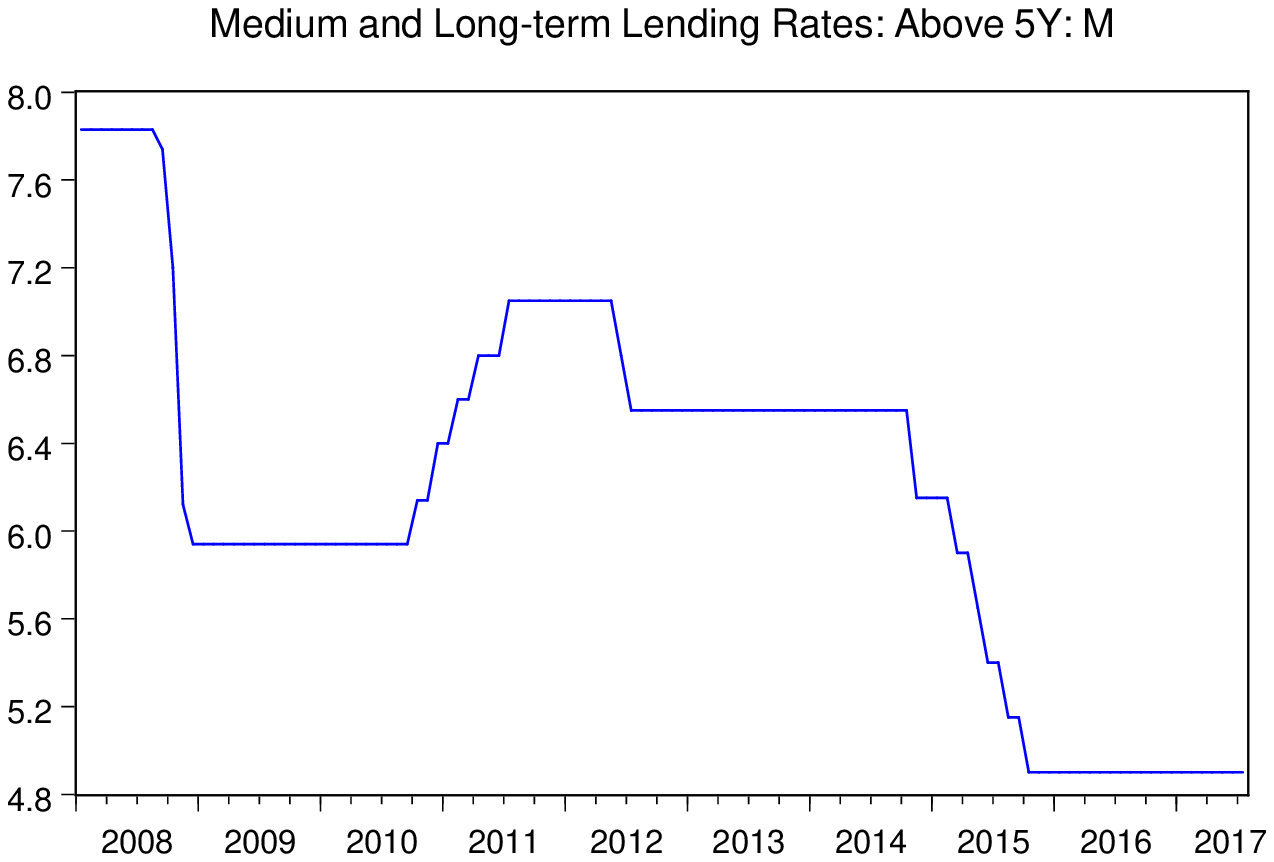}&\includegraphics[scale=0.45]{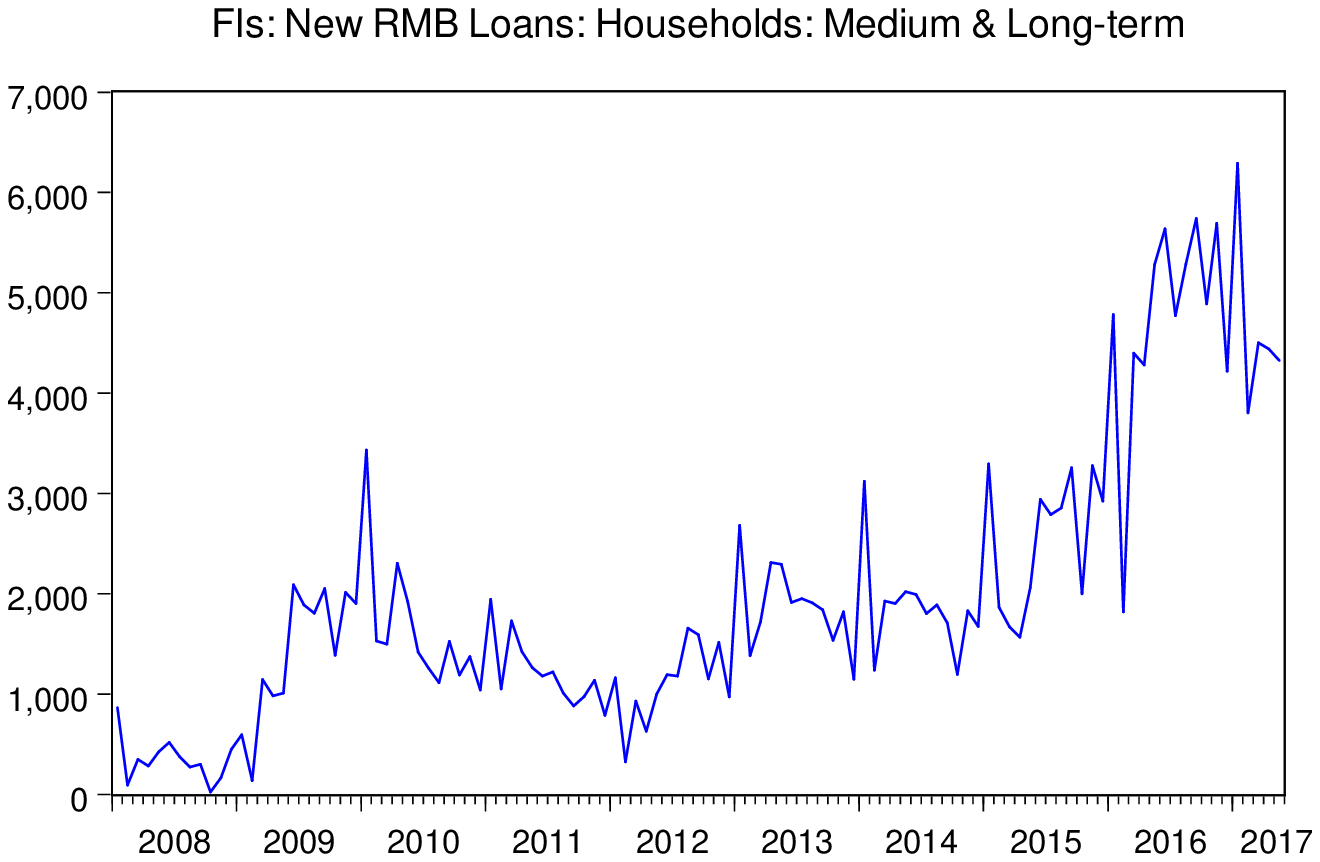}
    \\
    \includegraphics[scale=0.45]{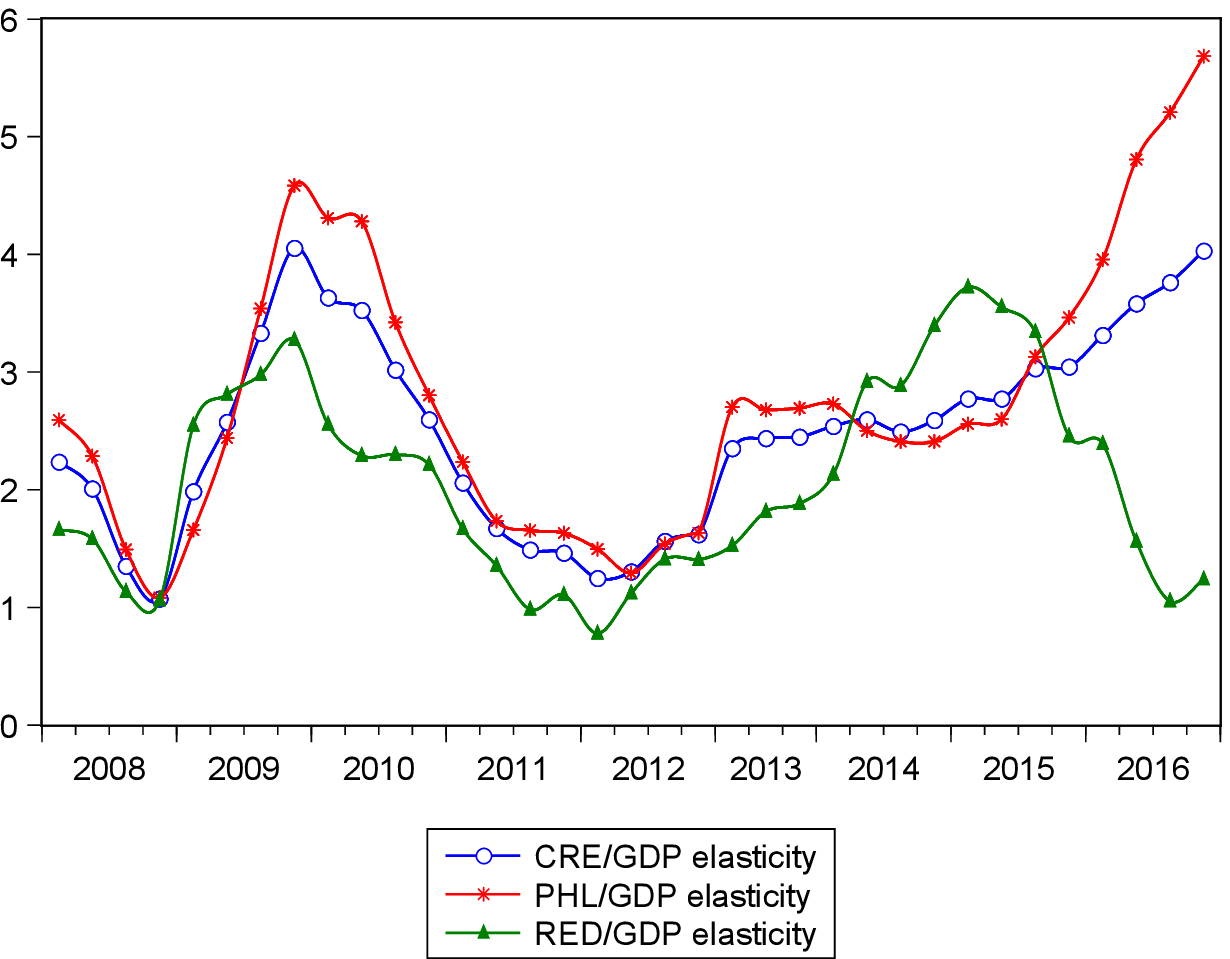}&\includegraphics[scale=0.45]{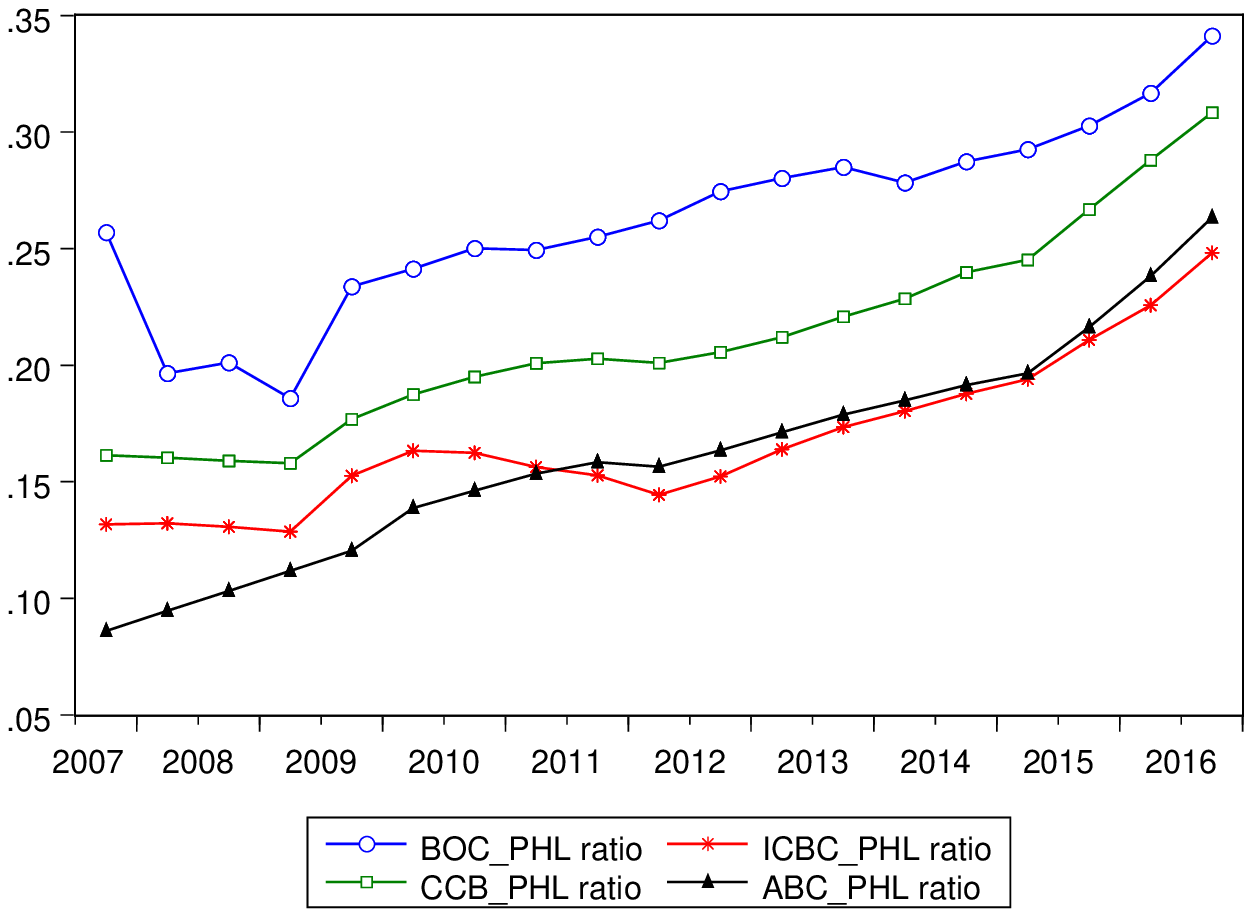}
  \end{tabular}
    \label{fig:credit}
\end{figure}

In order to gain an insight on the recent housing boom in China, one needs to trace back to the aftermath of 2007 Global Financial Crisis (GFC). In November 2008, the Chinese government had announced a 4-trillion investment package that aims to stimulate domestic economy\footnote{The package aims to assist ten key sectors, encompassing housing, rural infrastructure, transportation, health and education, environment, industry, disaster rebuilding, income-building, tax cuts, and finance. It was further announced that the central government provides only a quarter of the total package, with the rest three quarters being allocated by the budgets of the local and provincial governments through bank lending (citation).}. Despite a brief period of accelerating growth, the real sector had continued to slow down according to Fig. \ref{fig:GDP}. The slowdown is particularly noticeable in the Li Keqiang index - an alternative measurement of macroeconomic performance\footnote{The Li Keqiang index was coined by \emph{The Economist} magazine to measure China's economic performance in terms of three aspects:  the railway cargo volume, electricity consumption and loans disbursed by banks. It is widely cited as an alternative economic barometer compared to the official GDP figure \citep{E2010}.}. On the other hand, the credit sector had become increasingly lenient toward household borrowing. According to Fig. \ref{fig:credit}, we observe a rapid growth in medium-to-long-term household borrowing, with a steady decrease of interest rates (upper panel). There is also a dramatic rise in terms of the elasticity of Personal Housing Loan (PHL) relative to GDP\footnote{The elasticity is defined as $PHL\% / GDP\%$, which measures the percentage change of PHL relative to $1\%$ growth of GDP. Similarly, we also calculate the elasticity of Real Estate Development (RED) Loans as well as the Commercial Real Estate (CRE) loans relative to GDP in Fig. \ref{fig:credit}.}, as well as the proportion of PHL relative to the total amount of loans in the Big-4 Chinese banks\footnote{The Big-4 Chinese banks are Bank of China (BOC), China Construction Bank (CCB), Industrial and Commercial Bank of China (ICBC), and Agricultural Bank of China (ABC).} (lower panel). The traditional cultural urge to hold real estate, accompanied by a lack of alternative investment channels since the stock market crash in 2008 and 2015, had prompted households to participate even more aggressively in the spending spree over housing while giving banks further incentives to lend to households. The increasing level of financial fragility associated with the over-heating housing market had raised concerns of the central government. In the official newspaper \emph{People's daily (Ren Min Ri Bao)}, the potential danger of a real estate bust was, for the first time, referred to as a \emph{``grey rhino''}\footnote{See \citet{W2016}: \emph{The Gray Rhino: How to Recognize and Act on the Obvious Dangers We Ignore}.}, alluding to extreme events in large probability instead of a lower-probability Talebic \emph{``black swan''} \citep{T2007}. A series of purchasing restrictions was issued as early as 2010 in order to curb the over-heating market\footnote{the Guo-Shi-Tiao was announced and implemented in Tier-1 cities in Apr/2010, stating that one family can only purchase one apartment. The policy was further extended to other major Chinese cities in the subsequent five years.}. \citet{LPTW2014} conduct an experimental study on the efficacy of such asset holding caps and argue that such policy would be effective in eliminating bubbles if it is properly designed.

%

Unlike other breeds of speculative bubbles, the burst of a real-estate bubble could potentially engender far-reaching consequence due to the deep involvement of banking sector. Historically, the burst of Japanese real-estate bubble in the late 1980s had led the country into a decades-long recession that hasn't yet recovered to date. The full ramification of the latest sub-prime mortgage crisis in the US and the subsequent global contagion is still yet to be seen. In order to avoid systemic financial crisis and the ensuing economic depression, it is vital to conduct \emph{ex-ante} bubble diagnosis and provide well-informed policy guidance that aims to curb bubbles at its formative stage prior to the critical point.

As is pointed out by \citet{CS2003}, the term ``bubble'' is widely used but rarely clearly defined. Amongst existing studies, bubble is commonly referred to as a temporary deviation of price from its fundamental value. There are generally two strands of literature in the study of housing bubbles from such a fundamental approach. The first strand applies the idea of user-cost model of \citet{P1984}, which usually takes housing rent as a proxy of future cash flows and derives the equilibrium price of real estates based on non-arbitrage condition and efficient market hypothesis \citep{G2006,HKL2012}. The second strand studies the long-term relationship between housing prices and the underlying economic fundamentals. A cointegration model is generally applied and bubble is identified if there is no cointegration between housing prices and fundamental factors \citep{HY2006, MZ2007, AN2008}.

Yet the fundamental definition of bubble does not capture the inherent non-linear dynamics of bubble formation process that drives the bubble.  From a Minskyan point of view, bubbles are inherently unstable due to bounded rationality of investors \citep{M1975}. \citet{Kind1989} anatomized a typical financial bubble in five distinct phases: displacement, credit expansion, euphoria, overtrading, and revulsion. Hence, in addition to the fundamental definition of bubbles, we cannot overlook the speculative nature of a bubble driven by herd behaviour and aggravated by drastic credit expansion and self-reinforcing positive feedback mechanism. Such unsustainable price trajectory is often characterized by a faster-than-exponential growth in asset price during the bubble formation process prior to its burst. This is mathematically formalized by \citet{S2003} in the development of Log-Periodic-Power-Law-Singularity (LPPLS) model. The LPPLS framework had made a remarkable track record in early diagnosis of bubbles as well as post-mortem analysis of historical bubbles. \citet{SZ2004} studies the 2000 new economy bubble. \citet{ZS2006} examines the US Real-Estate Bubble. \citet{JZSWBC2010} conducted a bubble diagnosis of 2005-2007 and 2008-2009 Chinese stock market bubbles. More recently, \citet{ZSZ2017} investigate the critical transitions of Chinese real estate bubbles based on the LPPLS framework. From a laboratory setting, \citet{HMS2005} conduct a controlled price formation experiment, which detects the existence of super-exponential bubble due to positive feedback.

This paper aims to conduct a series of bubble diagnostic analysis over nine representative Chinese cities from two aspects. \emph{First}, we investigate whether the prices had been significantly deviating from economic fundamentals by applying a standard Engle-Granger cointegration test. \emph{Second}, we apply the Log-Periodic-Power-Law-Singularity (LPPLS) model to detect whether there is any evidence of unsustainable, self-reinforcing speculative behaviours amongst the univariate price series. We differ from \citet{ZSZ2017} in two aspects. \emph{First}, a conventional cointegration analysis is incorporated in comparison to the LPPLS analysis. \emph{Second}, we focus on diagnosing the LPPLS signals based on the filter condition instead of the critical time ($t_c$). On the policy side, we argue that it is vital to conduct bubble diagnostic test in order to identify whether bubbles are fundamental-driven or speculative-driven, and design policies that address the specific bubble signatures.

The rest of the paper is organized as follows. Section \ref{metho} outlines the methodological framework of bubble diagnosis from both fundamental and LPPLS perspectives. Section \ref{data} describes the data and defines the variables used in this study. Section \ref{res} provides the empirical results. Finally, section \ref{conclusion} concludes.

\section{Methodology}
\label{metho}
\subsection{The Fundamental Diagnosis of Bubbles: Pairwise Engle-Granger Approach}
We apply a simple pairwise cointegration model of \citet{EG1987} to investigate whether there is any long-run relationship between housing prices and its corresponding fundamental factors. Suppose there are two non-stationary series $X_t$ and $Y_t$ that are integrated of order one $I(1)$. If certain linear combination of $X_t$ and $Y_t$ yields I(0), then we say that $X_t$ and $Y_t$ are cointegrated. The Engle-Granger cointegration test is simply done in two steps: \emph{first}, we conduct the Augmented Dicky-Fuller (ADF) test\footnote{See \citet{DF1979}.} to see if all series are $I(1)$. \emph{Second}, we run a simple OLS regression of $Y_t=\beta_0+\beta_1 t + \beta_2 X_t+u_t$ and test if $u_t$ is $I(0)$. In the context of this paper, a \emph{fundamental bubble} exists if \emph{(i)} the price series are not $I(0)$, and \emph{(ii)} the fundamental factors are $I(0)$, \textbf{or} the $I(1)$ fundamental factor is not cointegrated with $I(1)$ price series.

\subsection{The Log-Periodic-Power-Law-Singularity (LPPLS) Diagnosis of Bubbles}
We further apply the LPPLS framework proposed by \citet{S2003} in order to examine whether the housing prices contain any sign of unsustainable speculative component characterized by faster-than-exponential growth pattern due to herd-behaviour and positive feedback mechanism, coupled with a periodic oscillation dynamics due to heterogeneous interactions between fundamentalists and trend chasers. Hence we apply the Johansen-Ledoit-Sornette (JLS) model, of which the law of motion takes the following form:
\begin{equation}
\frac{dp}{p}=\mu(t) dt +\sigma(t) dW - \kappa dj,
\label{eq:dp}
\end{equation}

\noindent where $\mu(t)$ is the drift and $\sigma(t)$ is the diffusion component of a standard Wiener process augmented by a discontinuous jump $dj$ with the dummy switch $j=0$ prior to the critical time and $j=1$ after the crash. We further define the crash hazard rate $h(t)$, which is proportional to the expectation of $dj$: $h(t)=E[dj]/dt$. By applying the rational expectation condition of Blanchard and Watson (1982), we obtain
\begin{equation}
\mu(t)=\kappa h(t).
\label{eq:mu}
\end{equation}

According to \citet{S2003}, the hazard rate takes the following form:
\begin{equation}
h(t)=\alpha (t_c - t)^{m-1}(1+\beta cos(\omega \ln(t_c -t) - \phi'),
\label{eq:h}
\end{equation}

Eq. (\ref{eq:h}) consists of a component of power law singularity due to self-reinforcing herd behaviour of noise traders $\alpha (t_c - t)^{m-1}$, as well as an oscillating component $\beta \cos(\omega \ln (t_c-t) - \psi')$ driven by the heterogeneous interactions between fundamentalists and trend chasers. The power-law singularity captures the positive feedback mechanism driven by the self-reinforcing beliefs while the log-periodicity embodies the tug-of-war between the two types of investors leading to large scale volatility leading to the singularity, or critical time ($t_c$). Combining (\ref{eq:dp} - \ref{eq:h}) yields
\begin{equation}
E[lnp(t)]=A + B(t_c - t)^m + C(t_c - t)^m \cos[\omega \ln(t_c - t) - \phi],
\label{eq:lnpt}
\end{equation}

\noindent with $B = - \kappa \alpha /m$ and $C=- \kappa \alpha \beta /\sqrt{m^2 + \omega ^2}$. Generally, the bubble regimes are conditioned by $0<m<1$ and $B<0$, where $m<1$ insures the singularity - the expected log-price diverges at $t_c$ and the two conditions combined insure that the price grows super-exponentially towards the critical time $t_c$.

Given its non-linear structure, the conventional method of calibrating Equation (\ref{eq:lnpt}) follows a two-step process. The first step involves a taboo search of Cvijovic and Klinowski (1995) to find N number of candidate solutions from our given search space. Each of these solutions is used as an initial estimates in a Levenberg-Marquardt non-linear least squares algorithm. We take the solution with the minimum sum of squares of the residuals. This paper adopts the procedure proposed by Filimonov and Sornette (2013) and Sornette et al. (2015) where the number of non-linear parameters has been reduced to just 3: $t_c$, $m$, $\omega$ for parsimonious purpose. The filter condition is given by Sornette et al. (2015) below:
\begin{eqnarray}
0.01 &\leq& m \leq 0.99
\\
2 &\leq& \omega \leq 25
\\
t_2 - 0.05 dt &\leq& t_c \leq t_2 + 0.1 dt
\\
2.5 &\leq& \frac{\omega}{2\pi}  \ln[\frac{t_c-t_1}{t_c-t_2}]
\\
1 &\leq& |mB/\omega C|
\end{eqnarray}

In the context of this paper, if all filter conditions above are met, we define a bubble in the corresponding series. We set the start time at $t_1=200801$, then move forward on a monthly basis until $t_1=201010$. The end time $t_2$ fixed at $t_2 = 201705$.

\section{Data Description}
\label{data}
Our data is collected from WIND and National Bureau of Statistic from Jan-2008 to Jun-2017 on a monthly basis\footnote{A database that provides comprehensive information on Chinese economy. URL: http://www.wind.com.cn/en/default.html}. We have chosen nine representative cities, which include the four Tier-1 cities, as well as five Tier 2\&3 cities: Beijing, Shanghai, Guangzhou, Shenzhen, Tianjin, Hangzhou, Chengdu, Wuxi, and Zhengzhou. We take the average price of traded residential building ($Yuan/m^2$) as a proxy for housing market condition ($P_{city}$). In addition, regional GDP ($GDP_{city}$), disposable income of urban residents ($INC_{city}$), and urban population of the respective cities ($POP_{city}$) are chosen as a measurement of fundamentals\footnote{The data is rearranged into monthly frequency by WIND. Monthly X-12 algorithm is also applied to eliminate seasonal effects in the series}. Furthermore, in order to analyse how credit factors influences the housing price dynamics, we take provident fund\footnote{It is a type of investment fund (especially in SE Asia) contributed to by employees, employers, and (sometimes) the state, out of which a lump sum is provided to each employee on retirement.} interest rate of individual housing loans ($IRH$), broad money ($M2$), as well as medium \& long-term new RMB loans ($NRLH$) granted to households ($>5 years$). The descriptive statistics of housing prices for Tier-1 cities and the representative Tier-2\&3 cities are presented in Table \ref{tab:DS1} and Table \ref{tab:DS23}.

\begin{table}[]
\centering
\footnotesize
\caption{Descriptive statistics of residential housing prices ($Yuan/m^2$) in Tier-1 cities: Jan/2008-Jun/2017. \emph{Source:} WIND Database}
\label{tab:DS1}
\begin{tabular}{lllll}
\hline
Price Statistic   & Beijing  & Shanghai  & Guangzhou & Shenzhen \\
             \hline
Mean         & 23234.07 & 21165.75  & 13604.13  & 25953.03 \\
Median       & 22323.50 & 17089.00  & 13916.00  & 20865.00 \\
Maximum      & 46834.00 & 50401.00  & 17882.00  & 61756.00 \\
Minimum      & 10486.38 & 7510.870  & 7977.821  & 10770.00 \\
\hline
\end{tabular}
\end{table}

\begin{table}[]
\centering
\caption{Descriptive statistics of residential housing prices ($Yuan/m^2$) for 5 representative cities outside Tier-1 category: Jan/2008-Jun/2017 \emph{Source:} WIND Database}
\footnotesize
\label{tab:DS23}
\begin{tabular}{llllll}
\hline
Price Statistic      & Tianjin & Hangzhou  & Chengdu     & Wuxi &  Zhengzhou        \\
\hline
Mean         & 10199.36 & 17012.28 & 8870.815 & 7729.790  & 6853.389 \\
Median       & 9998.000 & 16289.50 & 8881.000 & 7853.558  & 6789.000 \\
Maximum      & 17366.00 & 28214.00 & 14447.00 & 10591.94  & 10538.00 \\
Minimum      & 6470.760 & 11382.54 & 5336.630 & 4668.800  & 3771.000 \\ \hline
\end{tabular}
\end{table}
%
%
%

\section{Results}
\label{res}
\subsection{Are housing prices deviating from economic fundamentals in the long run?}
In this subsection, we examine whether housing prices are significantly derailed from the selected fundamental factors. We include three macro-variables on a regional basis, i.e. GDP, income per capita, and population of each representative cities. We first run a standard ADF test with constant and trend to examine whether the prices series and fundamental factors are stationary. If any price series is stationary, it indicates a self-correcting property in the series, hence no bubble. Table \ref{fundcoint1} reports the main result of our analysis. As for the price series, we found that all the prices are integrated of order 1 except Guangzhou. As for the macro-variables, we found that all the regional GDP series are non-stationary, yet for income per capita (\emph{INC}) and population (\emph{POP}), \emph{INC} of Hangzhou, Wuxi, and Zhengzhou, as well as \emph{POP} of Beijing and Wuxi are integrated of order zero ($I(0)$). We further conduct the Engle-Granger test to analyze whether the $I(1)$ price series are cointegrated with the $I(1)$ fundamentals. Generally, there are no clear sign of such cointegration except for Shanghai, where GDP is cointegrated with price at $5\%$ LOS and INC at $10\%$ LOS. Yet there is no evidence of cointegration between $P_{Shanghai}$ and $POP_{Shanghai}$. Overall, from this preliminary analysis, we conclude that only Guangzhou is excluded from the bubble hypothesis from the fundamental perspective.

\begin{table}[]
\centering
\caption{Engle-Granger diagnosis with fundamental factors. We denote residential housing price ($Yuan/m^2$) of city $x$ as $P_x$, per capita income of city $x$ as $INC_x$, and urban population of city $x$ as $POP_x$. It is found that $P_{Guangzhou}$, $POP_{Beijing}$, $INC_{Hangzhou}$, $INC_{Wuxi}$, $POP_{Wuxi}$, and $INC_{Zhengzhou}$ are $I(0)$ while the rest of variables $I(1)$. Guangzhou is excluded from the test since $P_{Guangzhou}$ is stationary. For other cities, a bubble exists since there is at least one fundamental factor that is not cointegrated with price series.}
\footnotesize
\label{fundcoint1}
\begin{tabular}{lllll}
\hline
Cities   & GDP      & INC      & POP      &  \\
\hline
Beijing & -1.54091 & -1.55562 & N/A &  \\
   & (0.8076)   & (0.8021)   & N/A     &  \\
Shanghai & -3.64674 & -3.26836 & -1.73829 &  \\
    & $(0.0317)^{\star\star}$   & $(0.0786)^{\star}$   & (0.7256)   &  \\
Guangzhou & N/A      & N/A      & N/A      &  \\
   & N/A      & N/A      & N/A      &  \\
Shenzhen & -2.23465 & -0.13013 & -1.3957  &  \\
   & (0.4644)   & (0.9936)   & (0.8555)   & \\
   \hline
Tianjin     & -2.08396 & -3.43469 & -1.84866 &  \\
             & (0.547)    & (0.0535)   & (0.6721)   &  \\
Hangzhou     & -2.72684 & N/A & -2.81171 &  \\
       & (0.2288)   & N/A    & (0.1973)   &  \\
Chengdu     & -1.39695 & -1.48165 & -1.32539 &  \\
        & (0.8552)   & (0.8285)   & (0.8749)   &  \\
Wuxi    & -0.91553 & N/A & N/A &  \\
        & (0.9489)   & N/A   & N/A    &  \\
Zhengzhou     & -2.59795 & N/A & -3.06301 &  \\
        & (0.2824)   & N/A   & (0.1219)   & \\
       \hline
\end{tabular}
\end{table}

\subsection{Is there unsustainable speculation in the housing market?}
As pointed out by \citet{Kind1989}, the bounded rationality of investors plays a pivotal role in the formation of speculative bubble. Furthermore, the endogenous nature of credit expansion renders such leveraged speculation possible. Such unsustainable, self-reinforcing bubble is often associated with an inherent relationship between asset prices and the underlying financial factors, and characterized by a faster-than-exponential growth component of price series, which is mathematically formalized by \citet{S2003} in the development of Log-Periodic-Power-Law-Singularity (LPPLS) model.

We begin with investigating whether there is any long-run relationship between housing prices and the underlying credit factors in a similar pairwise cointegration model. Three financial variables are selected, i.e. new RMB loans to households over five-years term ($NRLH$), provident fund interest rates of households ($IRH$), and broad money ($M2$). From the ADF tests, we found that all the three factors are $I(1)$. We further conduct a EG test between housing prices and the three financial factors, as is shown in Table \ref{tab:EG2}. The result is generally in line with \citet{XC2012}. It is evident that bank lending to household is strongly associated with the long-term surge of housing prices amongst most cities, as reflected in the cointegration between $P_x$ and $NRLH$ in the sample. The relationship is relatively weaker in $M2$ and $IRH$.
\begin{table}[]
\centering
\caption{Pairwise Engle-Granger test for real estate prices against financial variables. We denote New RMB Loans for Households as $NRLH$, Interest Rate for Households as $IRH$, and Broad Money as $M2$. There is a long-run relationship between housing prices and the underlying credit factors in Beijing, Shanghai, Shenzhen, Tianjin, Hangzhou, and Zhengzhou. The evidence is particularly strong in terms of $NRLH$.}
\label{tab:EG2}
\footnotesize
\begin{tabular}{lllllll}
\hline
Cities & $NRLH$ &  $IRH$ & $M2$\\
\hline
Beijing & $-2.782645$  & $-1.261070$  &  $-1.400137$ \\
   & $(0.06501)^{\star}$  & $(0.6443)$  &  $(0.9794)$ \\
Shanghai & $-2.939178$  & $-2.143475$ &  $-3.021259$ \\
   & $(0.0451)^{\star\star}$  & $(0.2285)$  &  $(0.0369)^{\star\star}$ \\
Guangzhou & $N/A$  & $N/A$ &  $N/A$ \\
   & $N/A$ & $N/A$  &  $N/A$ \\
Shenzhen & $-4.528026$  & $-2.242361$ &  $-1.891156$ \\
   & $(0.0004)^{\star\star\star}$  & $(0.1932)$  &  $(0.3350)$ \\
   \hline
Tianjin & $-3.384025$  & $-2.353078$ &  $-3.208008$ \\
   & $(0.0142)^{\star\star}$  & $(0.1582)$  &  $(0.0229)^{\star\star}$ \\
Hangzhou & $-3.275892$  & $-3.241492$ &  $-2.762250$ \\
   & $(0.0191)^{\star\star}$  & $(0.0210)^{\star\star}$  &  $(0.0681)^{\star}$ \\
Chengdu & $-2.500844$  & $-1.859283$ &  $-1.813710$ \\
   & $(0.1189)$  & $(0.3499)$  &  $(0.3706)$ \\
Wuxi & $-1.327366$  & $-0.798439$ &  $-0.837302$ \\
   & $(0.6136)$  & $(0.8143)$  &  $(0.8031)$ \\
Zhengzhou & $-3.136173$  & $-3.173356$ &  $-3.209019$ \\
   & $(0.0276)^{\star\star}$  & $(0.0250)^{\star\star}$  &  $(0.0228)^{\star\star}$ \\
\hline
\end{tabular}
\end{table}

\begin{table}[]
\centering
\caption{Occurrence of Log-Periodic-Power-Law-Singularity (LPPLS) Signals. There is evidence of LPPLS signatures in Shanghai, Shenzhen, Tianjin, and Chengdu.}
\footnotesize
\label{tab:LPPLSSS}
\begin{tabular}{lllll}
\hline
         & LPPLS Diagnosis & LPPLS Strength \\
          \hline
Beijing   & N               & $0\%$       &  \\
Shanghai  & Y               & $100\%$       &  \\
Guangzhou & N               & $0\%$       &  \\
Shenzhen  & Y               & $5.89\%$       &  \\
Tianjin   & Y                 & $5.89\%$       &  \\
Hangzhou  & N               & $0\%$       &  \\
Chengdu   & Y               & $17.65\%$       &  \\
Wuxi      & N               & $0\%$       &  \\
Zhengzhou & N               & $0\%$       &  \\
\hline
\end{tabular}
\end{table}
We further apply the LPPLS framework to diagnose whether the housing prices contain faster-than-exponential components according to \citet{JZSWBC2010}, as shown in Table \ref{tab:LPPLSSS}. The LPPLS strength here is defined as the proportion of windows with LPPLS signature over the total number of windows\footnote{We fix $t_2=201705$ and vary $t_1$ on a monthly basis from $t_1=200801$ to $t_1'=201010$. Hence we have a total number of 34 windows.}. We found four cities that exhibit signs of faster-than-exponential growth component, i.e. Shanghai, Shenzhen, Tianjin, and Chengdu. The LPPLS strength is particularly strong in Shanghai, of which all time windows are diagnosed with positive LPPLS signals. Surprisingly, Chengdu has a moderate level of LPPLS signal, despite the absence of relationship between $P_{Chengdu}$ and the underlying financial factors. It indicates that the housing price in Chengdu is likely to be driven by non-financial speculative factors. Shenzhen and Tianjin had shared a similar degree of LPPLS strength, although the housing price in Shenzhen is much higher in absolute terms.

Overall, we conclude that amongst the nine cities being examined, only Guangzhou does not exhibit clear signs of bubble due to its stationary property and absence of LPPLS signature, as summarized in Table \ref{tab:summary}. The result is generally in line with \citet{ZSZ2017}, which diagnosed an imminent high risk of turning point or correction in Beijing, Tianjin and Chengdu in 2017 while Shanghai and Shenzhen are at high risk in 2018 in terms of the quantile-based DS-LPPLS Confidence indicator. Similarly, there is no evidence of bubble in Guangzhou according to this previous study.

\begin{table}[]
\centering
\caption{Summary of Results: Engle-Granger (EG) vs. Log-Periodic-Power-Law-Singularity (LPPLS) Diagnosis. Only Guangzhou does not exhibit a clear bubble signature amongst the nine cities being examined.}
\footnotesize
\label{tab:summary}
\begin{tabular}{lllll}
\hline
          & EG  & LPPLS  & Overall &  \\
          \hline
Beijing   & Y            & N               & Y       &  \\
Shanghai  & Y            & Y               & Y       &  \\
Guangzhou & N            & N               & N       &  \\
Shenzhen  & Y            & Y               & Y       &  \\
Tianjin   & Y            & Y               & Y       &  \\
Hangzhou  & Y            & N               & Y       &  \\
Chengdu   & Y            & Y               & Y       &  \\
Wuxi      & Y            & N               & Y       &  \\
Zhengzhou & Y            & N               & Y       &  \\
\hline
\end{tabular}
\end{table}

It is evident from the empirical analysis above that the purchase \& credit restrictions enforced since April 2010 had by and large failed to curb excessive growth of housing prices, despite the strenuous effort of the central government. This is in contrast with \citet{LPTW2014}, which argues from a laboratory setting that such asset holding caps could be effective if it is well designed. Yet since 2017, the Chinese authority had announced another series of unconventional policies. In early 2017, a new policy named selling restriction (Xian-Mai) was implemented in 14 cities, which prohibits real estate buyers and investors to sell within a period of time that ranges from 2 years to 5 years. This policy had provided an efficient means to drive away short-term, speculative investors meanwhile preventing large-scale stampede from real estate markets that would eventually result in a crash. We recommend that, in order to design effective policy, it is vital to conduct bubble diagnostic test and implement relevant policies in cities with specific bubble signatures. For cities with significant LPPLS signature, it is vital to design policies that aim to eliminate excessive expectations on housing prices beyond mere demand-side control. For example, the selling constraint mentioned above would effectively eliminate the short-term speculative investors, thus stabilizing the price. Yet for cities that are diagnosed with fundamental bubbles without LPPLS signature, it would be more appropriate to design supply-side policies that accommodate the demand pressure arising from economic fundamentals.

\section{Conclusion}\label{conclusion}
The Chinese housing market had sustained a rapid growth since 2008, which is in stark contrast with the gloomy outlook of global economy and the steady decline of domestic real sector since the aftermath of 2007 GFC. It had raised concerns of both academics and policy makers alike.

This paper aims to provide an informed test over the bubble hypothesis of Chinese housing market based on nine representative cities. Our study differs from previous ones in the sense that we incorporate not only a conventional bubble definition from a fundamental perspective by means of a cointegration model, but also from a behavioural perspective of the emerging LPPLS framework to detect whether there is any unsustainable traits in the price dynamics characterized by faster-than-exponential growth signals. We found that amongst the nine cities being examined, only Guangzhou does not exhibit a clear sign of bubble.

On the policy side, we suggest that the conventional purchase \& credit restrictions haven't had the desired effect in curbing excessive increase of housing prices amongst the sample being examined. Given the heterogeneity that exists amongst cities with distinct bubble signatures, it is vital to conduct bubble diagnostic tests and implement relevant policies toward the specific bubble characteristics, rather than enforcing one-that-fits-for-all type policy that does not take into account such heterogeneity. We propose that, for cities with significant LPPLS signature, it is vital to implement policies that aim to eliminate speculative demand. In cities without LPPLS signature, the supply-side policies would be more appropriate in addressing the pressure arising from fundamental demand. The implications of such bubble diagnostic test toward a wider range of cities, as well as the efficacy of current policies will be further investigated in future research.
\newpage

\noindent \textbf{Acknowledgements}
\\
This research is supported by National Natural Science Foundation of China (No. 71721001, 71231008 \& 71671191) and
 Young Academic Mentorship Project of Sun Yat-sen University (No. 14000-31610161). We are grateful for the generous comments and suggestions from participants at International Conference on Econophysics (ICE) 2017 Shanghai, Computational Economics and Finance Conference (CEF) 2017 New York, and Artificial Economics Conference (AE) 2017 Tianjin. The usual disclaimer applies.
\newpage
\newpage
\noindent \textbf{Reference}
\bibliographystyle{elsarticle-harv}
\bibliography{Bibliography}

\newpage
\appendix{Appendix: Selected LPPLS outputs for cities with positive signals}\footnote{LPPLS filter rule: $Ind_{LPPLS} =1$ if $0.01 \leq m \leq 0.99$, $2 \leq \omega \leq 25$, $t_2 - 0.05 dt \leq t_c \leq t_2 + 0.1 dt$, $2.5 \leq \frac{\omega}{2\pi}  \ln[\frac{t_c-t_1}{t_c-t_2}]$, and $1 \leq |mB/\omega C|$, indicating a bubble exists in the corresponding time series. Otherwise $Ind_{LPPLS} =0$, indicating no evidence of LPPLS bubble.}.

\begin{table}[h!]
\centering
\caption{LPPLS Output: Shanghai}
\scriptsize
\label{sh}
\begin{tabular}{lllllllll}
\hline
t1                                                            & tc     & m     & w     & A      & B      & C     & Bm/Cw & $Ind_{LPPLS}$ \\ \hline
200801                                                        & 201705 & 0.746 & 4.245 & 10.883 & -0.004 & 0.001 & 1.438   & 1         \\
200802                                                        & 201705 & 0.74  & 4.247 & 10.89  & -0.005 & 0.001 & 1.45    & 1         \\
200803                                                        & 201705 & 0.713 & 4.251 & 10.923 & -0.006 & 0.001 & 1.51    & 1         \\
200804                                                        & 201705 & 0.712 & 4.247 & 10.923 & -0.006 & 0.001 & 1.513   & 1         \\
200805                                                        & 201705 & 0.747 & 4.267 & 10.883 & -0.004 & 0.001 & 1.43    & 1         \\
200806                                                        & 201705 & 0.75  & 4.289 & 10.882 & -0.004 & 0.001 & 1.412   & 1         \\
200807                                                        & 201705 & 0.776 & 4.393 & 10.867 & -0.004 & 0     & 1.317   & 1         \\
200808                                                        & 201705 & 0.779 & 4.402 & 10.866 & -0.003 & 0     & 1.309   & 1         \\
200809                                                        & 201705 & 0.781 & 4.408 & 10.865 & -0.003 & 0     & 1.303   & 1         \\
200810                                                        & 201705 & 0.775 & 4.384 & 10.868 & -0.004 & 0     & 1.323   & 1         \\
200811                                                        & 201705 & 0.765 & 4.332 & 10.872 & -0.004 & 0.001 & 1.356   & 1         \\
200812                                                        & 201705 & 0.763 & 4.316 & 10.873 & -0.004 & 0.001 & 1.365   & 1         \\
200901                                                        & 201705 & 0.751 & 4.228 & 10.879 & -0.004 & 0.001 & 1.408   & 1         \\
200902                                                        & 201705 & 0.748 & 4.186 & 10.88  & -0.004 & 0.001 & 1.424   & 1         \\
200903                                                        & 201705 & 0.716 & 4.086 & 10.912 & -0.006 & 0.001 & 1.484   & 1         \\
200904                                                        & 201705 & 0.708 & 3.947 & 10.912 & -0.006 & 0.001 & 1.511   & 1         \\
200905                                                        & 201705 & 0.703 & 3.839 & 10.911 & -0.006 & 0.001 & 1.523   & 1         \\
200906                                                        & 201705 & 0.7   & 3.77  & 10.91  & -0.006 & 0.001 & 1.526   & 1         \\
200907                                                        & 201705 & 0.702 & 3.824 & 10.911 & -0.006 & 0.001 & 1.525   & 1         \\
200908                                                        & 201705 & 0.707 & 3.913 & 10.912 & -0.006 & 0.001 & 1.523   & 1         \\
200909                                                        & 201705 & 0.715 & 4.07  & 10.913 & -0.006 & 0.001 & 1.515   & 1         \\
200910                                                        & 201705 & 0.721 & 4.159 & 10.913 & -0.005 & 0.001 & 1.508   & 1         \\
200911                                                        & 201705 & 0.715 & 4.067 & 10.913 & -0.006 & 0.001 & 1.513   & 1         \\
200912                                                        & 201705 & 0.751 & 4.263 & 10.882 & -0.004 & 0.001 & 1.477   & 1         \\
201001                                                        & 201705 & 0.762 & 4.429 & 10.88  & -0.004 & 0     & 1.454   & 1         \\
201002                                                        & 201705 & 0.768 & 4.482 & 10.878 & -0.004 & 0     & 1.443   & 1         \\
201003                                                        & 201705 & 0.767 & 4.468 & 10.878 & -0.004 & 0     & 1.445   & 1         \\
201004                                                        & 201705 & 0.774 & 4.553 & 10.877 & -0.004 & 0     & 1.435   & 1         \\
201005                                                        & 201705 & 0.806 & 4.878 & 10.868 & -0.003 & 0     & 1.365   & 1         \\
201006                                                        & 201705 & 0.805 & 4.867 & 10.869 & -0.003 & 0     & 1.368   & 1         \\
201007                                                        & 201705 & 0.793 & 4.756 & 10.872 & -0.003 & 0     & 1.389   & 1         \\
201008                                                        & 201705 & 0.768 & 4.499 & 10.877 & -0.004 & 0     & 1.393   & 1         \\
201009                                                        & 201705 & 0.763 & 4.447 & 10.878 & -0.004 & 0     & 1.386   & 1         \\
201010                                                        & 201705 & 0.785 & 4.669 & 10.874 & -0.003 & 0     & 1.413   & 1 \\ \hline
\end{tabular}
\end{table}

\begin{table}[]
\centering
\caption{LPPLS Output: Shenzhen}
\scriptsize
\label{bj}
\begin{tabular}{lllllllll}
\hline
t1                                                            & tc     & m     & w     & A      & B      & C     & Bm/Cw & $Ind_{LPPLS}$ \\
\hline
200801 & 201709 & 0.799 & 3.172 & 11.224   & -0.003  & 0.001 & 1.054 & 1 \\
200802 & 201709 & 0.879 & 3.142 & 11.172   & -0.002  & 0     & 1.038 & 1 \\
200803 & 201709 & 1     & 3.089 & 11.112   & -0.001  & 0     & 1.024 & 0 \\
200804 & 201709 & 1     & 3.126 & 11.114   & -0.001  & 0     & 1.011 & 0 \\
200805 & 201709 & 1     & 3.136 & 11.115   & -0.001  & 0     & 1.007 & 0 \\
200806 & 201709 & 1     & 3.126 & 11.114   & -0.001  & 0     & 1.011 & 0 \\
200807 & 201709 & 1     & 3.138 & 11.115   & -0.001  & 0     & 1.007 & 0 \\
200808 & 201709 & 1     & 3.196 & 11.118   & -0.001  & 0     & 0.988 & 0 \\
200809 & 201709 & 1     & 3.232 & 11.12    & -0.001  & 0     & 0.977 & 0 \\
200810 & 201709 & 1     & 3.237 & 11.121   & -0.001  & 0     & 0.975 & 0 \\
200811 & 201709 & 1     & 3.241 & 11.121   & -0.001  & 0     & 0.974 & 0 \\
200812 & 201709 & 1     & 3.254 & 11.122   & -0.001  & 0     & 0.971 & 0 \\
200901 & 201709 & 1     & 3.215 & 11.119   & -0.001  & 0     & 0.98  & 0 \\
200902 & 201709 & 1     & 3.173 & 11.116   & -0.001  & 0     & 0.99  & 0 \\
200903 & 201709 & 1     & 3.098 & 11.111   & -0.001  & 0     & 1.003 & 0 \\
200904 & 201708 & 1     & 2.946 & 11.097   & -0.001  & 0     & 1.019 & 0 \\
200905 & 201707 & 1     & 2.769 & 11.078   & -0.001  & 0     & 1.032 & 0 \\
200906 & 201707 & 0     & 2.192 & 166070.8 & -166058 & 0.262 & 0.626 & 0 \\
200907 & 201707 & 0     & 2.233 & 155980.8 & -155968 & 0.265 & 0.608 & 0 \\
200908 & 201707 & 0     & 2.261 & 151409.9 & -151397 & 0.27  & 0.588 & 0 \\
200909 & 201707 & 0     & 2.265 & 217427   & -217414 & 0.271 & 0.585 & 0 \\
200910 & 201707 & 0     & 2.276 & 72515.84 & -72503  & 0.273 & 0.578 & 0 \\
200911 & 201707 & 0     & 2.27  & 118581.4 & -118569 & 0.272 & 0.581 & 0 \\
200912 & 201707 & 0     & 2.275 & 100781.6 & -100769 & 0.273 & 0.579 & 0 \\
201001 & 201707 & 0     & 2.267 & 189005.6 & -188993 & 0.272 & 0.582 & 0 \\
201002 & 201707 & 0     & 2.243 & 118439.3 & -118426 & 0.271 & 0.588 & 0 \\
201003 & 201707 & 0     & 2.203 & 188795.2 & -188782 & 0.27  & 0.595 & 0 \\
201004 & 201707 & 0     & 2.192 & 105828.1 & -105815 & 0.27  & 0.596 & 0 \\
201005 & 201707 & 0     & 2.178 & 132157.1 & -132144 & 0.27  & 0.597 & 0 \\
201006 & 201707 & 0     & 2.174 & 176746.1 & -176733 & 0.27  & 0.597 & 0 \\
201007 & 201707 & 0     & 2.213 & 255119.7 & -255107 & 0.269 & 0.596 & 0 \\
201008 & 201707 & 0     & 2.233 & 162061.7 & -162049 & 0.269 & 0.594 & 0 \\
201009 & 201707 & 0     & 2.235 & 300033.9 & -300021 & 0.269 & 0.594 & 0 \\
201010 & 201707 & 0     & 2.225 & 125547.9 & -125535 & 0.269 & 0.595 & 0
\\ \hline
\end{tabular}
\end{table}

\begin{table}[]
\centering
\caption{LPPLS Output: Tianjin}
\scriptsize
\label{tj}
\begin{tabular}{lllllllll}
\hline
t1                                                            & tc     & m     & w     & A      & B      & C     & Bm/Cw & $Ind_{LPPLS}$ \\
\hline
200801 & 201706 & 0.659 & 2.227 & 9.797    & -0.005   & 0.001 & 1.29  & 1 \\
200802 & 201706 & 0.71  & 2.121 & 9.777    & -0.004   & 0.001 & 1.292 & 1 \\
200803 & 201705 & 0.801 & 1.953 & 9.753    & -0.002   & 0.001 & 1.296 & 0 \\
200804 & 201705 & 0.991 & 1.658 & 9.727    & -0.001   & 0     & 1.327 & 0 \\
200805 & 201705 & 1     & 1.677 & 9.727    & 0        & 0     & 1.309 & 0 \\
200806 & 201705 & 1     & 1.727 & 9.728    & 0        & 0     & 1.278 & 0 \\
200807 & 201705 & 1     & 1.769 & 9.729    & 0        & 0     & 1.254 & 0 \\
200808 & 201705 & 1     & 1.789 & 9.729    & 0        & 0     & 1.244 & 0 \\
200809 & 201705 & 1     & 1.815 & 9.73     & 0        & 0     & 1.23  & 0 \\
200810 & 201705 & 1     & 1.833 & 9.731    & 0        & 0     & 1.221 & 0 \\
200811 & 201705 & 1     & 1.82  & 9.73     & 0        & 0     & 1.228 & 0 \\
200812 & 201705 & 1     & 1.81  & 9.73     & 0        & 0     & 1.232 & 0 \\
200901 & 201705 & 1     & 1.815 & 9.73     & 0        & 0     & 1.23  & 0 \\
200902 & 201705 & 1     & 1.815 & 9.73     & 0        & 0     & 1.23  & 0 \\
200903 & 201705 & 1     & 1.787 & 9.729    & 0        & 0     & 1.244 & 0 \\
200904 & 201705 & 1     & 1.738 & 9.728    & 0        & 0     & 1.269 & 0 \\
200905 & 201705 & 1     & 1.686 & 9.728    & 0        & 0     & 1.296 & 0 \\
200906 & 201705 & 1     & 1.641 & 9.727    & 0        & 0     & 1.321 & 0 \\
200907 & 201705 & 1     & 1.567 & 9.726    & 0        & 0     & 1.363 & 0 \\
200908 & 201705 & 1     & 1.55  & 9.726    & 0        & 0     & 1.373 & 0 \\
200909 & 201705 & 1     & 1.523 & 9.726    & 0        & 0     & 1.388 & 0 \\
200910 & 201705 & 0.882 & 1.633 & 9.736    & -0.001   & 0     & 1.389 & 0 \\
200911 & 201705 & 0.816 & 1.69  & 9.744    & -0.002   & 0.001 & 1.404 & 0 \\
200912 & 201705 & 0.719 & 1.767 & 9.761    & -0.003   & 0.001 & 1.45  & 0 \\
201001 & 201705 & 0.663 & 1.798 & 9.775    & -0.005   & 0.001 & 1.492 & 0 \\
201002 & 201705 & 0.681 & 1.793 & 9.77     & -0.004   & 0.001 & 1.476 & 0 \\
201003 & 201705 & 0.679 & 1.793 & 9.771    & -0.004   & 0.001 & 1.479 & 0 \\
201004 & 201705 & 0.707 & 1.783 & 9.764    & -0.004   & 0.001 & 1.456 & 0 \\
201005 & 201705 & 0.762 & 1.754 & 9.753    & -0.002   & 0.001 & 1.423 & 0 \\
201006 & 201705 & 0     & 0.698 & 47912.66 & -47903   & 0.237 & 0.203 & 0 \\
201007 & 201705 & 0     & 0.71  & 46551.48 & -46541.9 & 0.239 & 0.183 & 0 \\
201008 & 201705 & 0     & 0.708 & 23253.06 & -23243.4 & 0.239 & 0.186 & 0 \\
201009 & 201705 & 0     & 0.712 & 13937.11 & -13927.5 & 0.239 & 0.18  & 0 \\
201010 & 201705 & 0.759 & 1.776 & 9.755    & -0.002   & 0.001 & 1.422 & 0 \\ \hline
\end{tabular}
\end{table}

\begin{table}[]
\centering
\caption{LPPLS Output: Chengdu}
\scriptsize
\label{cd}
\begin{tabular}{lllllllll}
\hline
t1                                                            & tc     & m     & w     & A      & B      & C     & Bm/Cw & $Ind_{LPPLS}$ \\
\hline
200801 & 201709 & 0.92  & 10.553 & 9.483    & 0        & 0     & 0.419 & 0 \\
200802 & 201709 & 0.926 & 10.491 & 9.482    & 0        & 0     & 0.429 & 0 \\
200803 & 201706 & 0.61  & 2.451  & 9.648    & -0.007   & 0.002 & 1.036 & 1 \\
200804 & 201705 & 0.68  & 2.261  & 9.62     & -0.004   & 0.001 & 1.081 & 1 \\
200805 & 201705 & 0.757 & 2.077  & 9.599    & -0.002   & 0.001 & 1.126 & 1 \\
200806 & 201705 & 0.864 & 1.832  & 9.578    & -0.001   & 0     & 1.189 & 0 \\
200807 & 201705 & 0.916 & 1.737  & 9.575    & -0.001   & 0     & 1.218 & 0 \\
200808 & 201705 & 0.981 & 1.623  & 9.574    & -0.001   & 0     & 1.259 & 0 \\
200809 & 201705 & 1     & 1.604  & 9.575    & -0.001   & 0     & 1.258 & 0 \\
200810 & 201705 & 1     & 1.625  & 9.576    & -0.001   & 0     & 1.237 & 0 \\
200811 & 201705 & 1     & 1.642  & 9.577    & -0.001   & 0     & 1.221 & 0 \\
200812 & 201705 & 1     & 1.652  & 9.577    & -0.001   & 0     & 1.211 & 0 \\
200901 & 201705 & 1     & 1.658  & 9.577    & -0.001   & 0     & 1.206 & 0 \\
200902 & 201705 & 1     & 1.659  & 9.577    & -0.001   & 0     & 1.206 & 0 \\
200903 & 201705 & 1     & 1.65   & 9.577    & -0.001   & 0     & 1.213 & 0 \\
200904 & 201705 & 1     & 1.636  & 9.577    & -0.001   & 0     & 1.225 & 0 \\
200905 & 201705 & 1     & 1.617  & 9.576    & -0.001   & 0     & 1.241 & 0 \\
200906 & 201705 & 1     & 1.59   & 9.576    & -0.001   & 0     & 1.262 & 0 \\
200907 & 201705 & 1     & 1.567  & 9.576    & -0.001   & 0     & 1.28  & 0 \\
200908 & 201705 & 1     & 1.553  & 9.576    & -0.001   & 0     & 1.291 & 0 \\
200909 & 201705 & 0.953 & 1.603  & 9.576    & -0.001   & 0     & 1.282 & 0 \\
200910 & 201705 & 0.855 & 1.75   & 9.58     & -0.001   & 0.001 & 1.246 & 0 \\
200911 & 201705 & 0.699 & 2.033  & 9.61     & -0.004   & 0.001 & 1.196 & 1 \\
200912 & 201706 & 0.532 & 2.359  & 9.674    & -0.013   & 0.003 & 1.152 & 1 \\
201001 & 201706 & 0.333 & 2.798  & 9.871    & -0.07    & 0.008 & 1.085 & 1 \\
201002 & 201707 & 0.147 & 3.258  & 10.598   & -0.512   & 0.023 & 1.007 & 1 \\
201003 & 201708 & 0.024 & 3.602  & 17.37    & -6.913   & 0.049 & 0.955 & 0 \\
201004 & 201708 & 0     & 3.638  & 137541.3 & -137531  & 0.055 & 0.965 & 0 \\
201005 & 201708 & 0     & 3.611  & 67216.53 & -67206   & 0.053 & 0.993 & 0 \\
201006 & 201708 & 0     & 3.6    & 94994.89 & -94984.4 & 0.051 & 1.01  & 0 \\
201007 & 201708 & 0     & 3.576  & 309174.8 & -309164  & 0.049 & 1.047 & 0 \\
201008 & 201707 & 0     & 3.558  & 63710.24 & -63699.8 & 0.047 & 1.079 & 0 \\
201009 & 201705 & 0.077 & 7.497  & 11.155   & -1.157   & 0.029 & 0.406 & 0 \\
201010 & 201705 & 0.065 & 7.534  & 11.471   & -1.453   & 0.033 & 0.384 & 0 \\ \hline
\end{tabular}
\end{table}

\end{document}